\documentclass[a4paper,10pt]{article}
\input{epsf}

\title{Fractional counting of authorship to quantify scientific research output}
\author{Vincenzo Carbone \\
Dipartimento di Fisica, Universit\`a della Calabria, Cosenza, Italy \\
IPCF/CNR - Universit\'a della Calabria, Cosenza, Italy
}

\begin{document}

\maketitle

\begin{abstract}
We investigate the problem of counting co-authorhip in order to quantify the impact and relevance of scientific research output through normalized \textit{h-index} and \textit{g-index}. We use the papers whose authors belong to a subset of full professors of the Italian Settore Scientifico Disciplinare (SSD) FIS01 - Experimental Physics. In this SSD two populations, characterized by the number of co-authors of each paper, are roughly present. The total number of citations for each individuals, as well as their h-index and g-index, strongly depends on the average number of co-authors. We show that, in order to remove the dependence of the various indices on the two populations, the best way to define a fractional counting of autorship  is to divide the number of citations received by each paper by the square root of the number of co-authors. This allows us to obtain some information which can be used for a better understanding of the scientific knowledge made through the process of writing and publishing papers.
\end{abstract}

\newpage

The ensemble of papers published by a scientist at a given epoch, which have been cited by the scientific community, contain useful information on the impact and relevance of the research output of the individual. In 2005 J.E. Hirsch introduced the celebrated \textit{h-index} \cite{hirsch}, which would represents a measure of research achievement, and depends on both the number of a scientist’s publications, and their impact on his or her peers. Simply said, the h-index is the highest number of papers signed by a scientist, that have each received at least that number of citations. Thus, someone with and h-index ranked $H$ has published $H$ papers each had at least $H$ citations. The h-index represents a better measure with respect to other bibliometric parameters as counting total papers, which could reward those with many mediocre publications, whereas counting just highest-ranked papers may not recognize a large and consistent body of work during a scientific career.

The parameter immediately attracted lots of attention of the scientific world, policy makers and the public media. The growth of the number of papers on the h-index is spectacular, and it is practically impossible to present a complete reference list (e.g. \cite{reazione1,reazione2,reazione3,reazione4,reazione5,reazione6,reazione7}). Scientific news editors \cite{reazione1} enthusiastically received the new index, and researchers in various fields of science \cite{batista,iglesias,popov,last}, particularly in the bibliometric research community \cite{reazione2,reazione3} started follow-up work. The idea of ranking scientists by a fair measure stirred the fire, because such rankings could make election procedures of scientific academies more objective and transparent. 

Apart for the simple definition of the h-index, the conclusions of the Hirsch's paper \cite{hirsch}, which are based on analysis of real data, are very interesting. Hirsch showed that it is hard to inflate one’s own h-index for example by self-citation, because the parameter relies on how a body of work is received over time and it is very hard to manipulate an entire career. Hirsch suggests that after 20 years in research, an $H \simeq 20$ is a sign of success, and $H \simeq 40$ indicates outstanding scientists likely to be found only at the major research laboratories. An $H \simeq 12$ should be good enough to secure university tenure \cite{hirsch}. What is also interesting in the data analysis by Hirsch is the fact that applying the method to prominent physicists, it can be found that $84\%$ of Nobel prize winners have substantial h-indices $H \geq 30$, while prominent physicists have $H \geq 50$ \cite{hirsch}, thus indicating that Nobel prizes, or even a brilliant scientific career, do not originate in one stroke of luck but in a body of scientific work. 

Among other, one of the main and perhaps the only serious disadvantage of the h-index has been revealed by L. Egghe \cite{egghe1,egghe2}, who noted that the h-index is insensitive to one or several outstandingly high cited papers. Indeed, although highly cited papers are important for the determination of the value $H$ of the h-index, once such a highly cited paper is selected to belong to the top $H$ papers, its actual number of citations at any time is not used anymore. Once a paper is selected to the top group, the h-index calculated in subsequent years remains insensitive to the citation of this paper, whatever the number of subsequent citations. To overcome this disadvantage of the h-index while keeping its advantages, it has been introduced the \textit{g-index} \cite{egghe1,egghe2}. Note that by definition the papers on rank $1 ,\dots, H$ each have at least $H$ citations, and hence these papers have, togheter, at least $H^2$ citations. The parameter $G$ defined through the \textit{g-index} \cite{egghe1} is just the largest rank such that the first $G$ papers have, together, at least $G^2$ citations. Obviously $G \geq H$ in all cases.

Actually a scientific work is made in general by collaborations among two or more scientists, so that lots of attention has been given to coauthored papers, say papers signed by more than one author \cite{egghe3}. How are the credits of each authore counted ? In other words, does every author in a $n$-authored paper get a credit of $1$ (total counting) or does every author get a credit of $1/n$ (fractional counting)? In general fractional counting is preferred because this does not increase the total weight of a single paper. The same question has been posed by Hirsch \cite{hirsch} which states that \textit{\dots a scientist with a high $H$ achieved mostry through papers with many co-authors would be treated overly kindly by his or her $H$. Subfields with typically large collaborations (e.g. high-energy experiments) will exhibit larger $H$ values, and I suggest that in cases of large differences in the number of co-authors, it may be useful in comparing different individuals to normalize $H$ by a factor that reflects the average number of co-authors} \cite{hirsch}. Possible solutions range from the simple division of $H$ by the average number of researchers in the publications of the Hirsch core\cite{batista,egghe3}, to discount the h-index for career length, multi-authorship and self-citations \cite{burrel}, and to take into account the actual number of co-authors and the scientist’s relative position in the byline \cite{wan}. Even if all the above proposal present advantages and disadvantages, in this paper I investigate how the fractional counting of autorship should simply work on a real case.

\begin{figure}
\epsfxsize=8cm 
\centerline{\epsffile{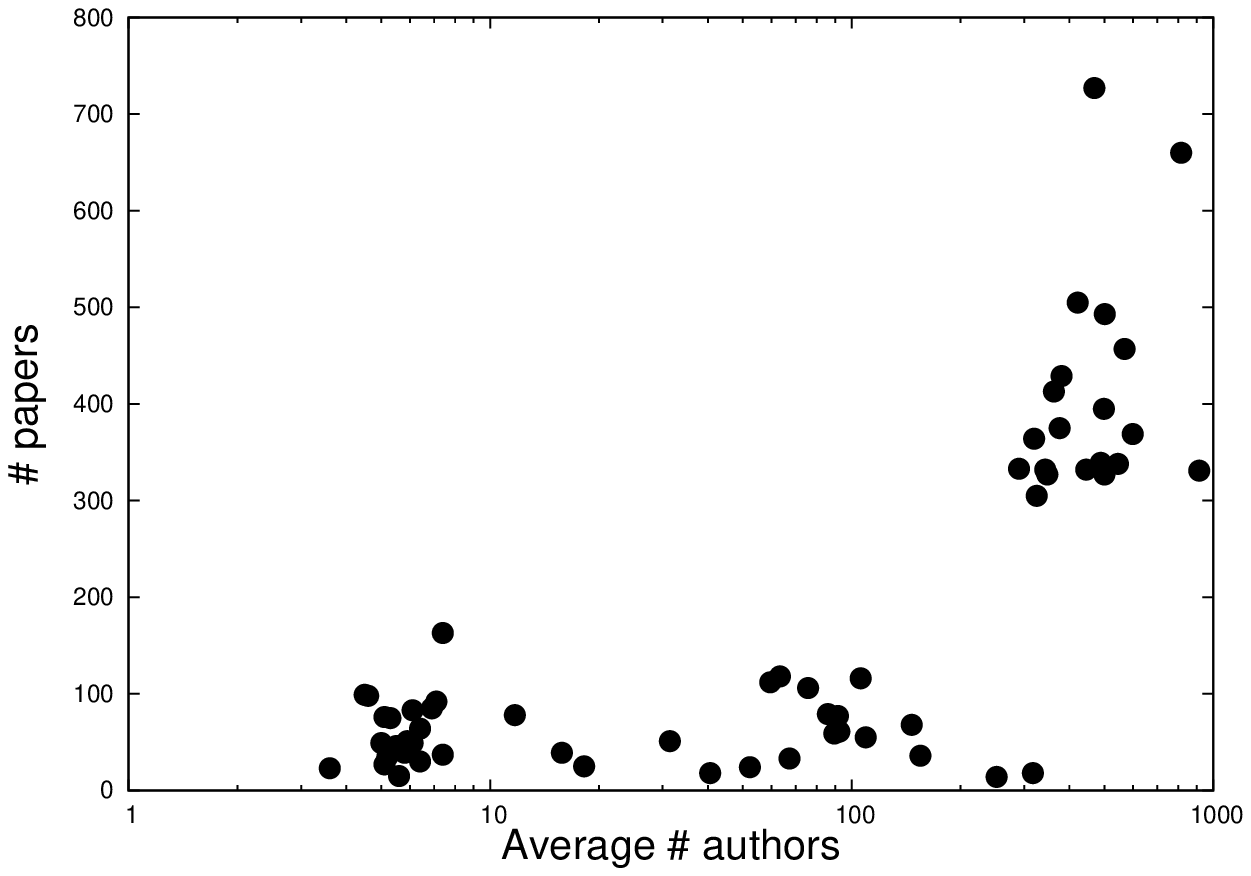}}
\epsfxsize=8cm 
\centerline{\epsffile{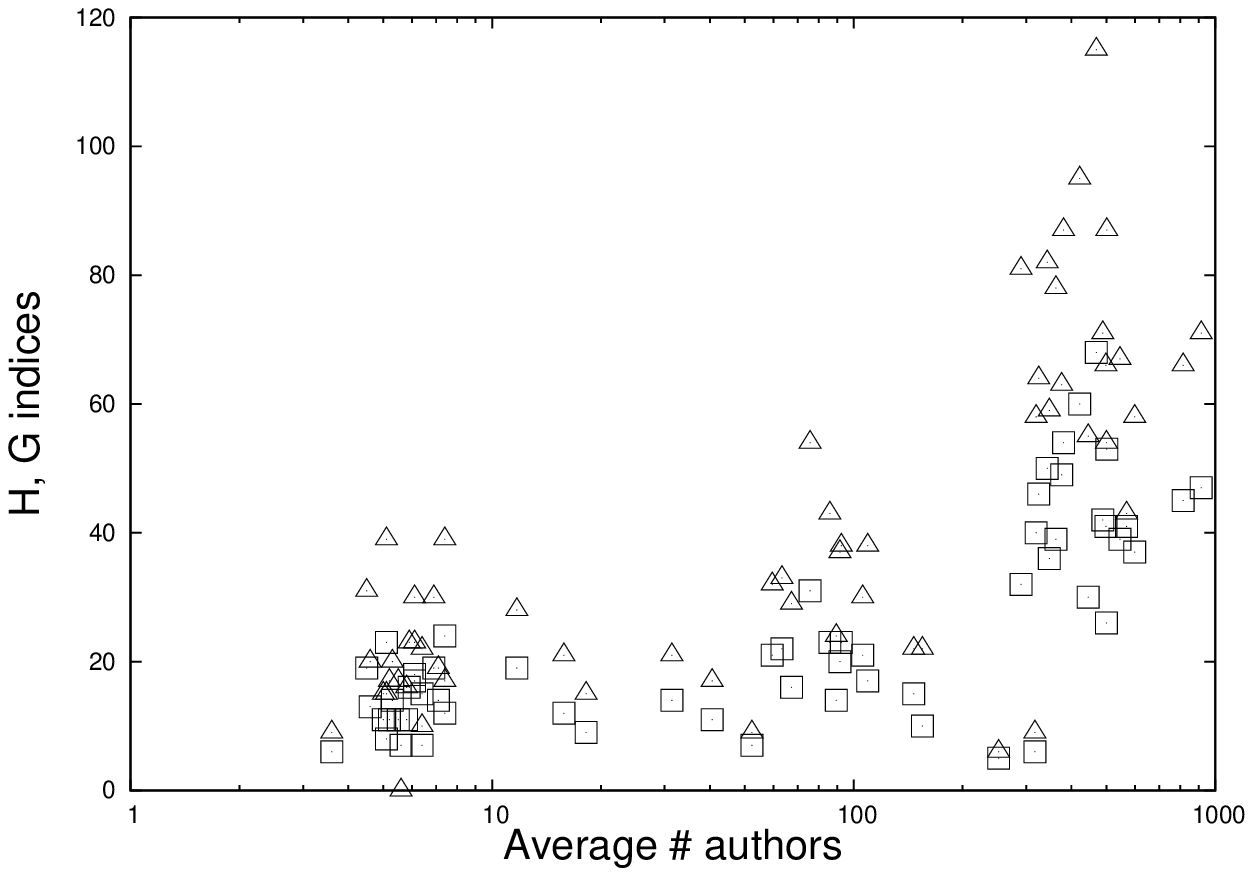}}
\caption{In the upper panel we report the number of papers as a function of the average number of coauthors for a given individuals. In the lower panel we report both the indices $H$ (squares) and $G$ (triangles) as a function of the average number of coauthors for a given individuals.}
\label{fig1}
\end{figure}

Aimed by a more accurate approach to fractional counting, we investigate scientific performances of a subset of anonymous individuals. We select individuals within the italian full professor of experimental physics belonging to the Settore Scientifico Disciplinare (SSD) FIS01. This choice is due to the fact that different individuals, belonging to uncomparable experimental facilities, coexist within this SSD. Using the Thomson ISI Web of Science database (available at http://isiknowledge.com), we select all the papers of a subset of $N=60$ full professors belonging to the above mentioned SSD, which roughly corresponds to $25\%$ of the whole full professors of the SSD. Let us consider, for each $j$-th individual ($j = 1,2, \dots, 60)$, the average value $M_j$ of authors of each publication $M_j$ and the total number of citations $C_{tot}$ of the $j$-th scientist at a given epoch. The value $M_j$ is correlated to the number of publications $n_j$, thus trivially both the usual h-index and g-index are correlated to $M_j$, as results from fig.s \ref{fig1}. As suggested by Hirsch \cite{hirsch}, the total number of citations is linearly related to $H^2$ through $C_{tot} = \alpha H^2$, with a parameter which results $\alpha = 4.45 \pm 0.06$, in agreement with Hirsch \cite{hirsch}. However, also the value of $G^2$ is related to $C_{tot}$ through the linear relation $C_{tot} = \beta G^2$, where $\beta = 1.68 \pm 0.02$ (not shown here).


What is interesting from fig.s \ref{fig1}, is that, as nively expected, two different populations are present within the SSD FIS01 which differ for the amount of the average number of co-authored papers. The two populations belong to the same SSD as it is well known, even if, as showed here, $n_j$ and both $H$ and $G$ are strongly dependent on $M_j$. In other words, the more the number of co-authors the higher the parameters which denote scientific performances. It goes without saying that it is much easier to get a high h-index when one has written many papers with many collaborators. Note that this is crucial if we conjecture that funding, tenure positions, etc. could be attributed on the basis of scientific performances. For example it could be conjectured that an individual might have an h-index greater than a threshold value $H \geq H_{th}$ in order to access a position of full professor in the SSD FIS01. It is clear that the non-homogeneity due to the two populations within FIS01 will make without sense an objective valutation. In order to avoid the rejection \textit{a priori} of the use of an objective scientometric index, we must investigate the problem of co-authorship.

I propose to weight each $i$-th paper of the $j$-th individual according to a fraction of the co-author number $m_i^{\mu}$ ($m_i$ is the number of co-authors of the $i$-th paper), and to compare the fractional indices which results from this operation. More formally, let us consider the weighted number of citation for each paper $\chi_{\mu}^{(i)} = C_i/m_i^\mu$, where $C_i$ is the number of citation collected by the $i$-th paper, ordered such as $\chi_{\mu}^{(1)} > \chi_{\mu}^{(2)} > \dots$. The fractional h-index $h_{\mu}$ and g-index $g_{\mu}$ for the $j$-th individual are then defined as the maximum integer such that 

\begin{equation}
\chi_{\mu}^{(h)} \geq h_{\mu}
\label{h}
\end{equation}
and 

\begin{equation}
\sqrt{\sum_{i=1}^{g_{\mu}} \chi_{\mu}^{(i)}} \geq g_{\mu}
\label{g}
\end{equation}

\begin{figure}
\epsfxsize=8cm 
\centerline{\epsffile{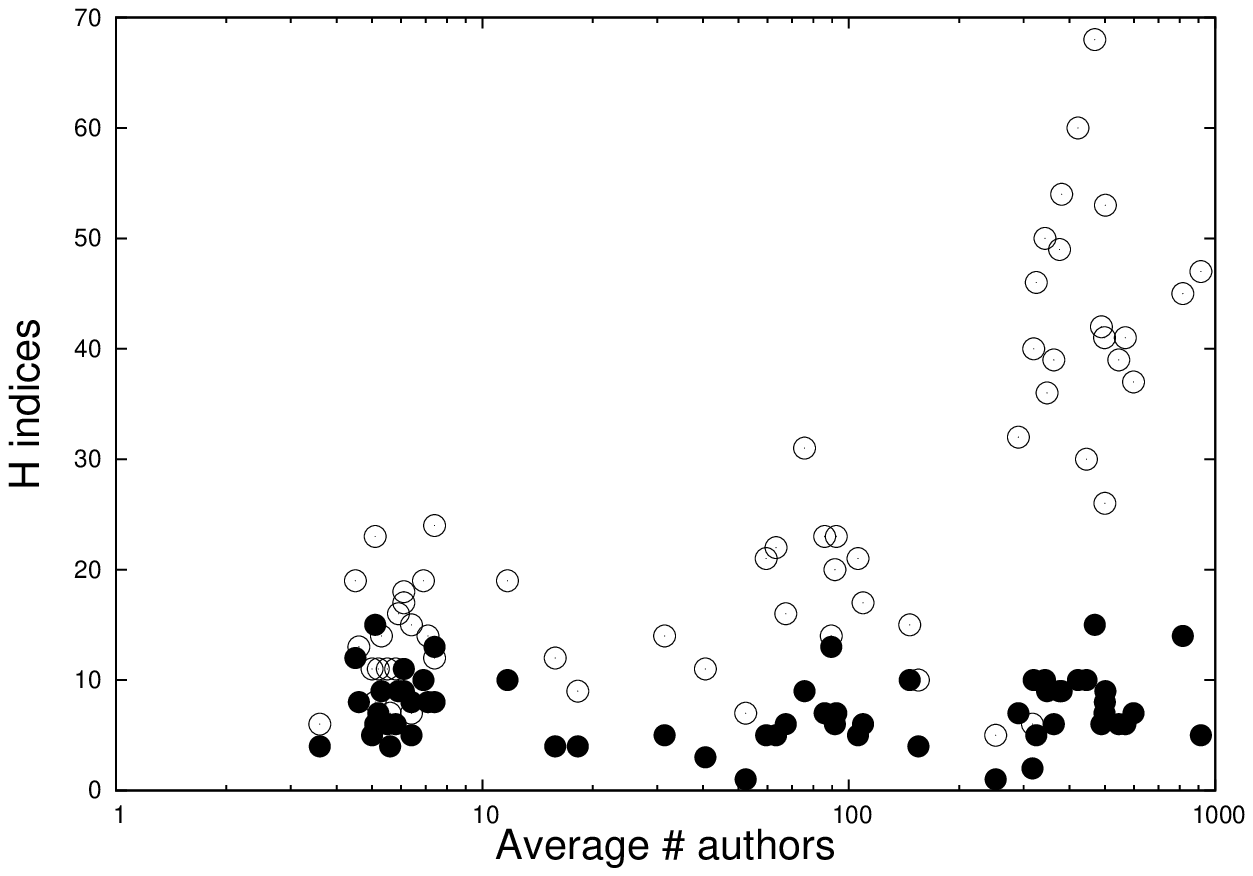}}
\epsfxsize=8cm 
\centerline{\epsffile{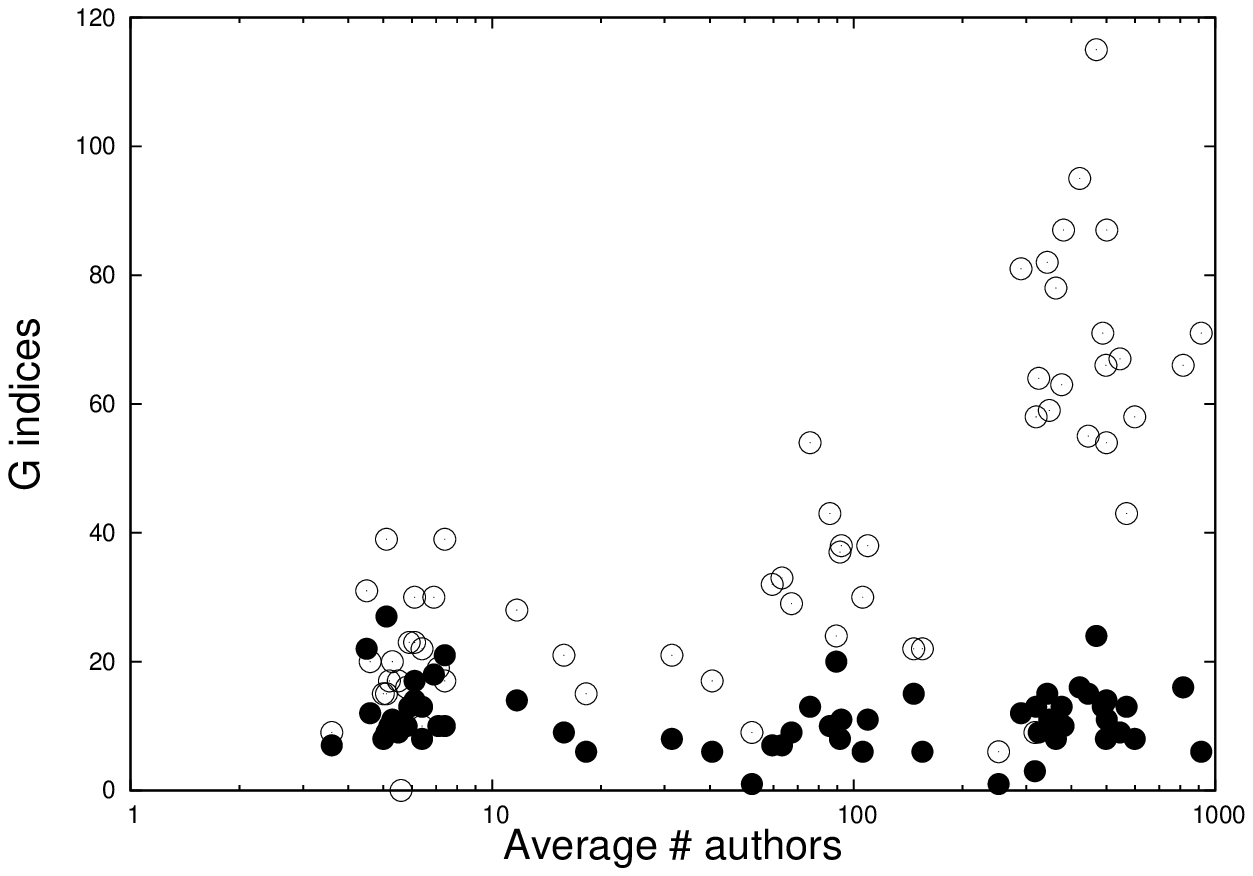}}
\caption{In the upper panel we report the values of both $H$ (withe symbols) and of their normalized value $h_{\mu}$ (black symbols) for $\mu = 0.53$. In the lower panel we report the values of both $G$ (withe symbols) and of their normalized value $g_{\mu}$ (black symbols) for $\mu = 0.53$.}
\label{fig2}
\end{figure}

A moment of reflection suffices to realize that the maximum weight $\mu = 1$ has the same effect of no weight $\mu = 0$, because the two popolation still should persist, even if with roughly upsetting their dependence on $M_j$. The most useful way to preceed is then to find the value of the parameter $0 < \mu < 1$ such that the resulting indices are independent on $M_j$. In other words we calculate a value $\mu^{\star}$ which minimizes the dependence of the parameters on $M_j$. Looking at fig.s \ref{fig1} we can conjecture that there roughly exists a linear relation between $h_{\mu}$ and $M_j$, and between $g_{\mu}$ and $M_j$. Then, by using the relation

\begin{equation}
f_p(N,\mu) = \left[ N \sum_{j=1}^N M_j p_{\mu} - \sum_{j=1}^N M_j \sum_{j=1}^N p_{\mu}\right]
\label{effe}
\end{equation}
(the index $p$ stands for both $h$ and $g$). we can simply define a best parameter through 

\begin{equation}
\mu^{\star}(N) \simeq \min_\mu \left\{f_g(N,\mu) + f_h(N,\mu)\right\}
\label{mustar}
\end{equation}
Using our dataset made by $N = 60$ individuals as an example, we find that the best-fit parameter which minimize the dependence on $M_j$ results $\mu^{\star} \simeq 0.53 \pm 0.01$, curiously close to $1/2$. In fig. \ref{fig2} we report the values of both $h_{\mu}$ and $g_{\mu}$ as a function of $M_j$, where the independence of both normalized indices on $M_j$ is clearly showed. It is worth reporting that also the sqyares of fractional indices are linearly related to the fractional numer $C_{tot}^{(\mu)}$ of total citation obtained by summing the fractional citations $\chi_{\mu}^{(i)}$ over all papers. This is made through the linear coefficients $C_{tot}^{(\mu)} = \alpha_\mu h_{\mu}^2 = \beta_\mu g_{\mu}^2$, where $\alpha_\mu = 5.45 \pm 0.08$ and $\beta_\mu = 2.04 \pm 0.07$.

\begin{figure}
\epsfxsize=8cm 
\centerline{\epsffile{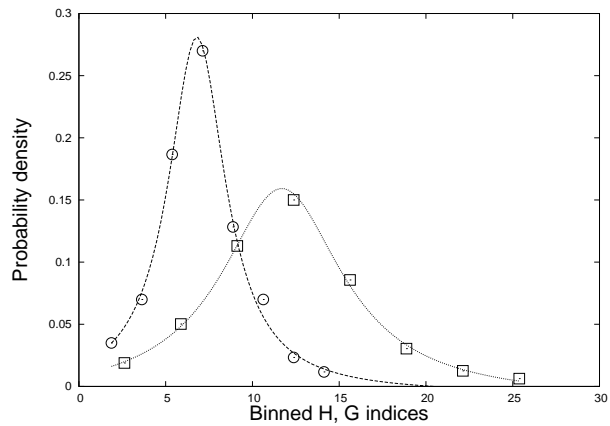}}
\caption{We report the binned values of both $h_{\mu}$ (circles) and $g_{\mu}$ (squares) for $\mu = 0.53$. Superimposed as full lines we report the fitted Cauchy-Lorentz functions $L_h(x)$ and $L_g(x)$ (see text).}
\label{fig3}
\end{figure}

The problem of the determination of typical values $h_{\mu}^{\star}$ and $g_{\mu}^{\star}$, and mainly their fluctuations within a given ensemble of individuals, should be of some practical interest. To evaluate these values we can build up the hystogram of both $h_{\mu}$ and $g_{\mu}$ calculated for $\mu = 0.53$, which are reproduced in fig. \ref{fig3}. The empirical distributions for both normalized indices can be very well reproduced through a Cauchy-Lorentz distribution function 

\begin{equation}
L_p(x) = L_0 + \frac{2A}{\pi} \frac{\sigma_p}{4(x-p)^2+\sigma_p^2}
\label{lorenziana}
\end{equation}
The maximum values $p$ of the distributions can be considered as typical h-index and g-index for the class of scientists at hand, while typical fluctuations are described by the values of $\sigma_p$. In our example, the best fit correspond to $h_{\mu}^{\star} \simeq 6.80 \pm 0.01$ with $\sigma_h \simeq 4.0 \pm 0.1$, and $g^\star \simeq 11.70 \pm 0.07$ with $\sigma_g \simeq 8.5 \pm 0.6$. 

The information we obtained can be used to infer something about scientific processes of knowledge. The fact that the best way to overcome the difficulty of co-authorship seems to be weighting each paper by the square root of the number of authors of that paper is quite evocative of a random walk dynamics. This should perhaps indicate that in big experimental collaborations, whose output is a paper with a lot of co-authors, the effective work is carried out independently by relatively small group of scientists, as usually happens in smaller laboratories within Universities. Moreover, the occurrence of a Cauchy-Lorentz distribution for normalized indices indicates that the various scientists tend to differentiate enough to generate a process of homogeneous broadening. Very interestingly, the fact that $\sigma_g > \sigma_h$ in the distribution functions means that the normalized g-index is the result of a larger broadening with respect to the h-index. This indicates that actually a succesfull scientific career is the result of some few research papers with a great impact and some more papers with fewer citations. 

In conclusion, I investigated the problem of how to weight a co-authored paper in order to not reject a priori the possibility of objectively using parameters as the h-index or the g-index. I introduced the fractional indices $h_{\mu}$ and $g_{\mu}$ built up by weighting the citations of the $i$-th paper with a power $\mu$ of the number of co-authors. The best fit parameter which minimize the strong dependence of the number of papers, citations and indices on the average number of co-authors if close to $\mu \simeq 1/2$. More interstingly, we found that, at least within the SSD FIS01 where two populations of scientists coexist, the above fractional count can gives rise to a single population. The information on the distribution functions of normalized indices could be very useful, for example, during the selection procedures of scientific academies, research funding and tenure decisions, which are often seen as opaque, clubby and capricious. In fact the hypothetical commettee could be free to use a threshold values $h_{th} \simeq h^\star - r \sigma_h$ and $g_{th} \simeq g^\star - r \sigma_g$, where $r$ is an arbitrary parameter, as one of the objective criteria to select younger scientists. Of course this is just one of the possible way to overcome the problem, and different methods can be investigated, even if they might be aimed at the solution of the presence of a double popolation. Bibliometric indicators as the normalized h-index and g-index, which as we showed are useful parameters to evaluate the output of science and which gives us some information about the way scientists actually work, cannot be considered as the only yardstick to evaluate the career of an individuals.

Acknowledgments: I'm very grateful to C. Basile and S. Donato who provided a partial dataset used in the paper. I'm grateful to P. Veltri and R. Bartolino for fruitful discussions.

\newpage

\end{document}